\documentclass{iopjournal}

\usepackage{amsmath} 
\usepackage{amssymb} 
\usepackage[
    backend=biber,
    sorting=none, 
    style=numeric-comp 
    ]{biblatex} 
\hypersetup{colorlinks=true, 
            linktoc=page}
\usepackage{cleveref} 
\usepackage{lmodern} 

\begin{document}
\title{Simulating dust collapse: the impact of gauge choices on primordial black hole formation} 

\author{Robyn L. Munoz$^{1,2*}$\orcid{0000-0003-3345-8520}, Christian T. Byrnes$^1$\orcid{0000-0003-2583-6536}, Marco Bruni$^{3,4}$\orcid{0000-0003-2305-5699} and Ian Hawke$^{5}$\orcid{0000-0003-4805-0309}}

\affil{$^1$ Department of Physics and Astronomy, University of Sussex, Brighton, BN1 9QH, United Kingdom}

\affil{$^2$ Advanced Research Computing at Cardiff, Redwood Building, King Edward VII Avenue, Cardiff, Wales, CF10 3NB}

\affil{$^3$ Institute of Cosmology \& Gravitation, University of Portsmouth, Dennis Sciama Building, Burnaby Road, Portsmouth PO1 3FX, United Kingdom}

\affil{$^4$ INFN Sezione di Trieste, Via Valerio 2, 34127 Trieste, Italy}

\affil{$^5$ Mathematical Sciences and STAG Research Centre, University of Southampton, Southampton SO17 1BJ, United Kingdom}

\affil{$^*$Author to whom any correspondence should be addressed.}

\email{MunozR2@cardiff.ac.uk}


\begin{abstract}

We investigate why numerical relativity simulations of dust collapse in a matter-dominated era frequently fail before primordial black hole formation due to numerical limitations and inappropriate gauge choices. Using 3-dimensional fully non-linear simulations of a dust fluid, we detail the advantages of advanced numerical methods and the impact of the lapse and shift evolution equations. The challenge is twofold: 1) conventional lapse slicing evolution equations, highly effective in typical black hole simulations, must be adapted in a cosmological context. We demonstrate how different proposed lapse evolution equations perform, enabling fair comparison, and propose our own. 2) The shift evolution equation’s role is to spread out the numerical grid observers for a stable simulation. However, the Gamma-driver evolution fails when confronted with shell-focussing of the dust fluid. We show how the shift damping choice depends on the resolution of the shell-focusing region, suggesting that future work should have spatially varying damping. Our improved gauge choices allow simulations to evolve significantly further and potentially reach black hole formation. The best-performing simulations reach close proximity to a black hole; however, further work on the gauge choice is required for this to be stable.
\end{abstract}

\section{Introduction} \label{sec:Intro}

In the classic 1975 paper \cite{Carr:1975qjz}, Carr compared the timescales of collapse vs the outward pressure perturbation, and estimated the collapse threshold for a primordial black hole (PBH) to match the equation of state parameter $\omega$: $\delta_c\simeq \omega$. This implies the probability of forming a PBH is enhanced for softer equations of state than radiation. Subsequent numerical \cite{Musco:2012au} and analytic treatments \cite{Harada:2013epa} have found consistent results. However, these estimates assume spherically symmetric overdensities, an approximation that works well during radiation domination but fails as $\omega\rightarrow0$. An early matter-dominated era, with $\omega=0$, is a well-motivated case to study in this limit \cite{Allahverdi:2020bys}\footnote{Note we are not considering a scalar field dominated era, e.g.~slow reheating \cite{Padilla:2024iyr}.} and one in which crucial deviations from spherical symmetry make analytic and numerical work much harder. Most studies, dating back to the 1980s \cite{Polnarev1981,Polnarev:1986bi}, used the probability distribution of deviations from spherical symmetry in terms of the shape of the initial overdensity \cite{Doroshkevich}, finding that the PBH fraction has a power-law dependence on the initial perturbation amplitude \cite{Harada:2016mhb,Kokubu:2018fxy,Harada:2022xjp}, rather than exponential, as is the case for radiation domination. There remains no consensus, with \cite{Ebrahimian:2025syf} concluding, based on analytic arguments, that PBH formation is only slightly more likely during matter domination than during radiation domination. Settling this question remains essential for understanding the phenomenology and constraints on an early matter-dominated era. 

In addition, from the perspective of PBHs adopted here, the study of gravitational collapse of inhomogeneities in a dust fluid is, at least formally, the same as the study focused on in the post-recombination matter era, when cold dark matter is the dominant component. Indeed, here restrictions apply only to the initial conditions: a super-horizon perturbation on an Einstein-de Sitter background, see \cref{sec:IC}, which is then evolved non-linearly in full general relativity. This can therefore be self-similarly transferred to either an early or late matter era, should observational constraints allow.

Our approach uses numerical relativity tools to simulate direct-collapse BH formation with pressureless matter described as a perfect fluid. This description of matter introduces a major numerical challenge: shell-crossing and shell-focusing \cite{Szekeres_1995}. Both describe scenarios in which nearby particles with opposing velocities, with no pressure to keep them apart, fly past each other and cause the simulation to fail. In this hydrodynamics context, we do not have particles but a fluid where this phenomenon induces a form of hypersurface folding onto itself. Shell-crossing occurs with shock fronts, resulting in a pancake-like singularity in the fluid\footnote{This is a type of singularity that is extendable \cite{Szekeres_1995}, the spacetime is locally Minkowski \cite{Wainwright_Ellis_1997}, and it is thought to correspond to the caustic momentarily forming in the case of particles, like in the Newtonian Zeldovich approximation.}, depending on the shear and the Weyl curvature \cite{Wainwright_Ellis_1997, Galoppo_2026}. Shell-focusing occurs where matter converges, which may be point-like (as in FLRW big-crunch) or cigar/spindle-like (see the Kasner solution for more on this \cite{Kasner_1921, Eardley_1971, Landau_1975, Wainwright_Ellis_1997}). With our focus on an inhomogeneity with three identical principal directions, simulations crash as they evolve toward a point-like shell-focusing singularity at the centre of the overdensity, as analysed in \cref{sec:RW}. 

The first simulations applied to matter BHs were spherically symmetric \cite{Harada_1998}, where BHs formed or not depending on the overdensity size and amplitude. This creates an interesting scenario to explore from the perspective of critical collapse, where the shell-focusing singularity is identified as a naked singularity \cite{Snajdr_2006, Gundlach_2007, Ortiz_2011, Gundlach_2025}. As PBH research has developed, analytical work has shown that certain initial conditions can lead to BH formation before shell-focusing \cite{Harada_2015, Galoppo_2026}, depending on the shape and amplitude of the overdensity, which identifies cases that can or cannot be simulated. As numerical tools have become more powerful and reliable, exploring matter PBH formation with 3-dimensional simulations has become possible \cite{deJong_2023, Yoo_2026}. However, the shell-focusing limitation has led to the use of dust-like matter \cite{deJong_2023}. And simulations where matter BHs are identified \cite{Yoo_2026} are short-lived, because of the shell-focusing crash that occurs shortly afterwards. Progress in this field would benefit from a shell-focussing singularity-avoiding gauge, so we endeavour to characterise the problem in this paper.

Building on previous work on matter-dominated numerical relativity simulations \cite{Munoz_2023}, we explore the PBH formation scenario by creating initial conditions in \cref{sec:IC}, implementing them in simulations in \cref{sec:NR}, and including additional tools to support stability in \cref{sec:Num}. With this setup, we can study the impact of the lapse and shift evolution equations in \cref{sec:Lapse,sec:Shift&MR}, then assess whether the BH criteria are met in \cref{sec:BHcrit}. Throughout this work, we use Greek indices for spacetime (0 to 3) and Roman indices for space (1 to 3), and we use units with $c=G=1$.

\section{Initial conditions} \label{sec:IC}

For this cosmological context, we start with a homogeneous FLRW metric with the dust Einstein-de Sitter solution, meaning a matter-dominated era with zero pressure. We include a perturbation in the initial conditions to allow an inhomogeneity to grow. Here it is implemented in the synchronous comoving gauge with our code \texttt{ICPertFLRW} \cite{ICPertFLRW, ICPertFLRW_GRH}, see the formalism described in \cite{Bruni_2014, Munoz_2023}, where the line element is
\begin{equation} \label{eq:ini_line_element}
    ds^2 = -d\tau^2 + \gamma_{ij} dx^i dx^j.
\end{equation}
The spatial metric $\gamma_{ij}$ is only given scalar perturbations, because vector and tensor perturbations are initially subdominant. Previous work \cite{Bruni_2014} has shown that these scalar perturbations can be directly expressed in terms of the gauge-invariant comoving curvature perturbation $\mathcal{R}_c$ \cite{Lyth_1985}, such that the spatial metric is \cite{Munoz_2023}:
\begin{equation} \label{eq:ini_metric}
    \gamma_{ij} = a^2 \left( 1 - 2 \mathcal{R}_{c} \right) \delta_{ij} - \frac{4}{5 H^2} \partial_{i}\partial_{j}\mathcal{R}_{c},
\end{equation}
where the scale factor $a$, the Hubble scalar $H = 2/3\tau$, and the background energy density $\bar{\rho}=3H^2/8\pi$ are defined by the Einstein-de Sitter solution.

The perturbation in the metric is analytically propagated throughout to get an exact initial extrinsic curvature $K_{ij} = -\frac{1}{2}\partial_{\tau}\gamma_{ij}$, and energy density $\rho$ from the Hamiltonian constraint equation \cite{Munoz_2023}.

One then has the freedom to define $\mathcal{R}_c$, as long as it is initially time independent. In line with previous work \cite{Munoz_2023}, considering symmetric boundary conditions, and focusing on a single cosmological over-density depicted by a sinusoidal mode, we choose:
\begin{equation} \label{eq:Rc}
    \mathcal{R}_c = \sum_{i = 1 \rightarrow 3} {
    A_{x^i}\cos \left( \frac{2\pi x^i}{L} \right)}.
\end{equation}
The perturbation's amplitude $A_{x^i}$ and wavelength $L$ are tuned such that the maximum density contrast $\delta = \rho / \bar{\rho} - 1$ is initially 0.03. Here $L$ is the comoving wavelength of the perturbation, which is chosen such that the physical initial scale is 4 times the initial Hubble radius $\lambda_{\rm phy, \; IN} = a_{\rm IN} L = 4/H_{\rm IN}$. This represents a large inhomogeneity, but allows the simulation to start linearly and let non-linearities grow rapidly, collapsing within a reasonable amount of computing time.

\begin{figure}[t]
    \centering
    \includegraphics[width=0.6\linewidth]{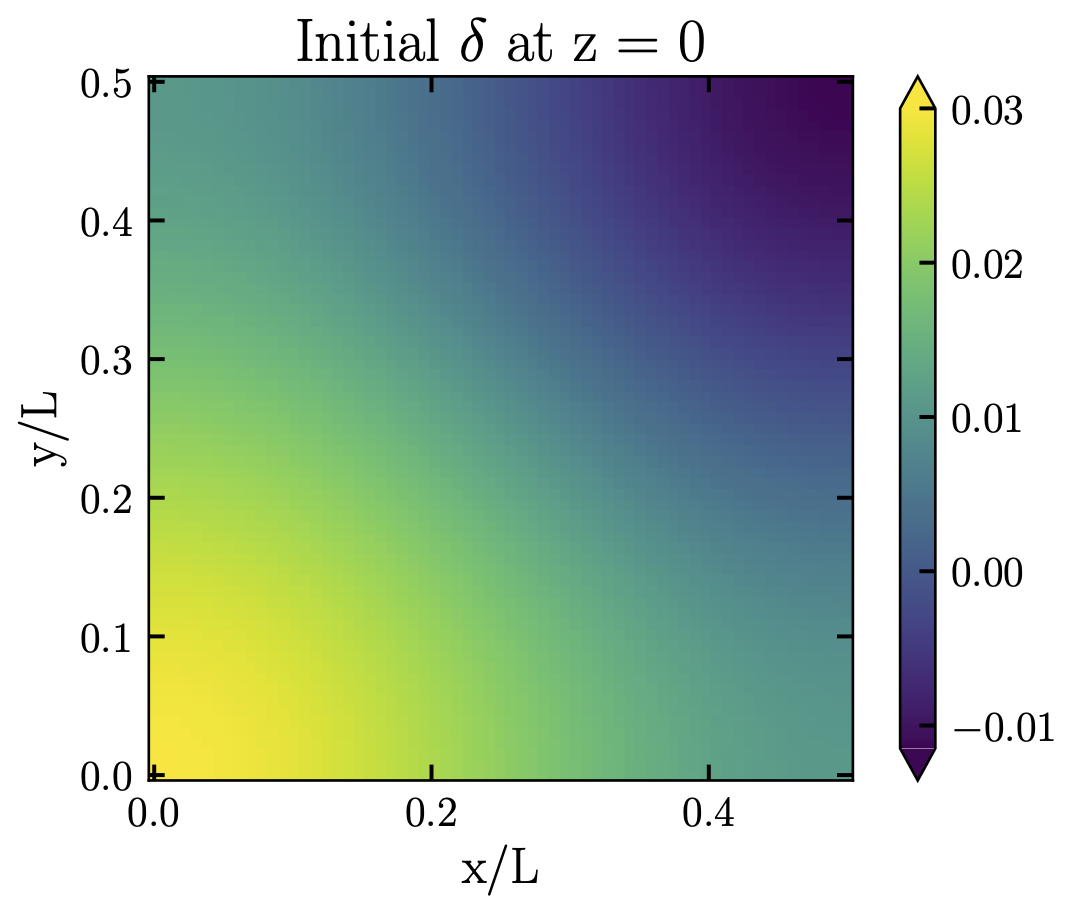}
    \caption{Initial density contrast $\delta = \rho / \bar{\rho} - 1$ in the $z=0$ plane of the simulation box, showing the sinusoidal profile with symmetric boundary conditions. The peak over-density is at the origin with $\max(\delta)=0.03$, while the bottom of the under-density is off this plane, at $x=y=z=0.5L$. Grid positions are at the centre of cubic cells.}
    \label{fig:ICdensity}
\end{figure}

The perturbation $\mathcal{R}_c$ in \cref{eq:Rc} is used to define the initial spacetime with \cref{eq:ini_line_element,eq:ini_metric}, which provide the energy density from the Hamiltonian constraint, fully defining the simulation's initial conditions. To visualise this distribution, see the density contrast in the $z=0$ plane in \cref{fig:ICdensity}. There are $64^3$ points on this coarse level, which we later refine with $8$ mesh refinement levels in \cref{sec:Shift&MR}, and the simulation box spans $[0, L/2]^3$ with symmetric boundary conditions to effectively simulate this full cosine grid, with the over-dense centre being at $x=y=z=0$.

\section{Numerical Relativity Simulation} \label{sec:NR}

From this starting point, this simulation is evolved fully non-linearly, with \texttt{Einstein Toolkit} \cite{ET2012, ET2025} following standard numerical relativity practices where the full spacetime is split according to the ADM formalism \cite{Alcubierre_2008}:
\begin{equation} \label{eq:sim_line_element}
    ds^2 = -(\alpha^2 + \beta_k\beta^k)dt^2 + 2\beta_idtdx^i + \gamma_{ij}dx^idx^j.
\end{equation}
The lapse $\alpha$ and shift $\beta^i$ are initially set to $\alpha=1$ and $\beta^i=0$, so that the line element matches \cref{eq:ini_line_element}, but their respective evolution equations are set according to the various gauge choices in \cref{sec:Lapse,sec:Shift&MR,sec:Num}, as provided by our \texttt{CosmoLapse} \cite{CosmoLapse, Kranc} code developed for this work.

The spatial metric $\gamma_{ij}$ is evolved according to the extrinsic curvature
\begin{equation}
    K_{ij} = -\frac{1}{2\alpha}\left(\partial_t\gamma_{ij} - \mathcal{L}_{\beta}\gamma_{ij}\right),
\end{equation}
with $\mathcal{L}_{\beta}$ the Lie derivative along the shift. In turn, the extrinsic curvature $K_{ij}$ is evolved according to Einstein's field equations, typically split into its trace $K=\gamma^{ij}K_{ij}$ and traceless $A_{ij}$ parts: 
\begin{equation}
    K_{ij} = \frac{1}{3}\gamma_{ij}K + A_{ij}.
\end{equation}

Einstein's field equations are formulated such that they are presented as stable 2nd-order partial differential equations via the $W$ BSSNOK formulation \cite{Campanelli_2005, Alcubierre_2008}. This takes a conformal transformation of the spatial metric and the traceless part of the extrinsic curvature
\begin{equation}
    \tilde{\gamma}_{ij} = W^{2}\gamma_{ij},
    \qquad
    \tilde{A}_{ij} = W^{2}A_{ij}.
\end{equation}

The spacetime is evolved by \texttt{McLachlan} \cite{McLachlan, Kranc} according to the evolution equations for $W$, $\tilde{\gamma}_{ij}$, $K$, $\tilde{A}_{ij}$ and most importantly $\tilde{\Gamma}^i = - \partial_{j}\tilde{\gamma}^{ij}$, the contracted Christoffel symbol of the conformal metric, which is treated within BSSNOK as an independent quantity.

For matter, the evolution method we use is dictated by the available hydrodynamics codes. We start with \texttt{CT\_Dust} \cite{CT_Dust} which uses the Wilson formulation \cite{Wilson_1972}. Then, when High-Resolution Shock Capturing (HRSC) is included, we use the \texttt{GRHydro}\footnote{As the \texttt{GRHydro} thorn doesn't currently support a cosmological equation of state, the pressureless matter was implemented in \texttt{EOS\_Omni} with a polytropic equation of state set to $p = 10^{-50} \rho^{2}$, making the pressure smaller than numerical noise in the simulation, and the atmospheric masks were turned off.} code \cite{GRHydro} based on the Valencia formulation \cite{Marti_1991}.

Having evolved all quantities with their allocated codes and evolution equations, we extract relevant diagnostics using our post-processing analysis package \texttt{aurel} \cite{aurel, EBWeyl} so that we can quantify the impact of various numerical and analytical choices in the following sections.

\subsection{Notion of time}

As we explore different gauges, each simulation will have different observers whose differing lapse and shift will induce a different notion of code time $t$, as introduced in \cref{eq:sim_line_element}. To make these comparable, we focus on the value of various observables as measured at the peak of the density distribution, located at the origin, where we calculate within our \texttt{CosmoLapse} code the observer's proper time $\tau$ from
\begin{equation} \label{eq:tau}
    d\tau = \sqrt{\alpha^2 - \beta_k\beta^k} \; dt. 
\end{equation}

With this, we draw comparisons with the spherical collapse model (based on a top-hat homogeneous over-density) \cite{Gunn_1972}. Indeed, previous work has shown that in the synchronous comoving gauge the peak evolves exactly as predicted by this model \cite{Munoz_2023}. It describes a spherical over-density whose radius expands with the expansion of the universe but eventually reaches its maximal size, then turns around and collapses in on itself. In doing so, it predicts specific values for the linear density contrast $\delta^{(1)}$, for the turnaround time $\delta^{(1)}_{TA}$, and collapse time $\delta^{(1)}_{C}$:
\begin{equation}
    \delta^{(1)}_{TA} = \tfrac{3}{20}(6\pi)^{2/3} \simeq 1.062,
     \hspace{2cm}
    \delta^{(1)}_{C} = 2^{2/3} \; \delta^{(1)}_{TA} \simeq 1.686.
\end{equation}
This comes from Einstein-de Sitter linear perturbation theory, which relates this linear density contrast $\delta^{(1)}$ to the proper time $\tau$ calculated with \cref{eq:tau} as
\begin{equation} \label{eq:delta1}
    \delta^{(1)} = \delta_{IN} \frac{a}{a_{IN}} = \delta_{IN} \left(\frac{\tau}{\tau_{IN}}\right)^{2/3}.
\end{equation}
Since $\delta^{(1)}$ reaches specific values at key moments of the evolution, we use this rescaling as the notion of time (instead of $\tau$) on the $x$-axis in the following figures, and mark with dashed vertical black lines the expected turnaround and collapse times.

However, once the shift is present, the volume element of the spatial metric no longer varies purely following gravitational in-fall, but also according to how the observer's positions are shifted by $\beta^i$. If nearby observers are shifted away from the centre, the volume element of the peak grid position will increase, directly affecting the density measurement. This is not accounted for in \cref{eq:delta1} as the larger area would need to be rescaled by a different initial density contrast $\delta_{IN}$. We can directly see that simulations evolving beyond $\delta^{(1)}_{C}$ are impacted by the shift's evolution.

\section{Numerical methods} \label{sec:Num}

Most problems in numerical relativity are linked to evolved variables covering too wide a range of scales to be computed with fixed-precision floating arithmetic. At best, the accuracy of the numerical simulation degrades to the point where the results are useless. At worst, the numerical values underflow or overflow, and the values can no longer be represented at all with finite precision. This is a ``\emph{crash}'': avoiding these issues is one of the field's central problems.

There are three standard ways in which these issues arise.

First, when evolving non-linear matter, the fields generically steepen. This can occur either through collapse (as in our case), where (for example) the density diverges, or through characteristic breaking and shock formation (a standard issue in all simulations of hydrodynamics). As gradients increase, simple numerical derivatives become prone to unphysical oscillations (Gibbs oscillations), leading to the over/underflow errors noted. Standard techniques, generally known as \emph{High-Resolution Shock Capturing} (HRSC) methods, have been developed to mitigate these problems \cite{LeVeque_2002}. In essence, HRSC methods apply a data-dependent dissipation term non-linearly to the evolved matter variables, ensuring the total variation (and hence the size of any oscillations) is bounded.

Secondly, the sub-principal terms in the evolution equations for the spacetime, combined with standard numerical techniques, may induce small oscillations in the spacetime variables. Left untreated, these small oscillations can grow similarly to Gibbs oscillations, producing the same over/underflow errors. Adding numerical Kreiss-Oliger dissipation \cite{Kreiss_1973} is the standard technique for damping the growth of these oscillations, and has been shown to have no significant impact on numerical accuracy.

Finally, the coordinate conditions, or gauge, chosen for the numerical evolution may either fail to cover the interesting parts of the spacetime, or may themselves lead to numerical instabilities. Here, the situation is the most complex, with no one ``standard technique'' that can be broadly applied. A ``good gauge'' must allow the evolution to cover as much of the spacetime as possible, but without inducing such steep gradients in the evolved variables as to induce a catastrophic loss of accuracy or a crash due to over/underflow. This is difficult for BHs, where the gauge must ensure the slice does not reach the physical singularity (singularity avoidance). A range of singularity-avoiding slicings have been proposed, some of which work in a broad range of situations for asymptotically flat spacetime evolutions \cite{Reimann_2003, Reimann_2004, Campanelli_2005}. However, when the generic behaviour of the spacetime is expanding (as in cosmological problems) whilst the gauge must also capture BH formation via collapse, there are no off-the-shelf solutions available.

To start, as a point of comparison, we consider the synchronous comoving gauge ($\partial_{t} \alpha = 0$ and $\partial_{t} \beta^i = 0$), matching previous work \cite{Munoz_2023}. This is the black line in \cref{fig:HRSC_diss}, which shows $\alpha=1$ throughout, and the density $\rho$ decreases with the dilution of the expanding universe. This over-dense region then decouples from this expansion and collapses inward, with density growing exponentially, and the simulation crashes as theoretically predicted by the spherical collapse model, marked by the vertical dashed black line. This crash results from a combination of the three effects described above. So we pre-emptively consider a different gauge, allowing us to appreciate the impact of certain numerical choices and start avoiding shell-focusing. We jump to the puncture gauge, the magenta curve in \cref{fig:HRSC_diss}, using the $1+\log$ cosmologically adapted slicing, and the Gamma-driver shift evolution equations:
\begin{equation} \label{eq:puncture}
\begin{aligned}
    \partial_{t}\alpha &= - 2 \alpha \; (K + 2/\tau), \\
    \partial^{2}_{t} \beta^{i} &= 0.75 \; \partial_{t}\tilde{\Gamma}^i - \partial_{t} \beta^{i},
\end{aligned}
\end{equation}
which are explained in more detail in \cref{sec:Lapse,sec:Shift&MR} respectively. Allowing the lapse and shift to evolve, we directly see that $\alpha \neq 1$ on the right hand panel of \cref{fig:HRSC_diss}. There are initial oscillations, followed by an evolution beyond the collapse time $\delta^{(1)}_{C}$. However, this is quickly dominated by finite difference error. This is hinted at in the discontinuity of the magenta curve, but is obvious upon visual inspection of the matter distribution in the simulation. We then successively implement numerical methods to stabilise this collapse for as long as possible.

\subsection{High-resolution shock capturing}

HRSC methods are included using the piecewise parabolic reconstruction method \cite{Colella_1984} and the HLLE Riemann solver \cite{Einfeldt_1988}, provided by \texttt{GRHydro} \cite{GRHydro}. Adding HRSC to a synchronous simulation makes it evolve according to the cyan dashed curve in \cref{fig:HRSC_diss}, where we find the same result, so the shell-focusing limitation of the synchronous gauge is indeed what makes the simulation crash. 

On the other hand, applying HRSC to a simulation with a puncture gauge immediately increases the simulation duration. Indeed, the red curve in \cref{fig:HRSC_diss} shows a stable decrease in the lapse, and as the shift increases the centre's volume, the density can stabilise.

\subsection{Dissipation}

\begin{figure}[t]
    \centering
    \includegraphics[width=\linewidth]{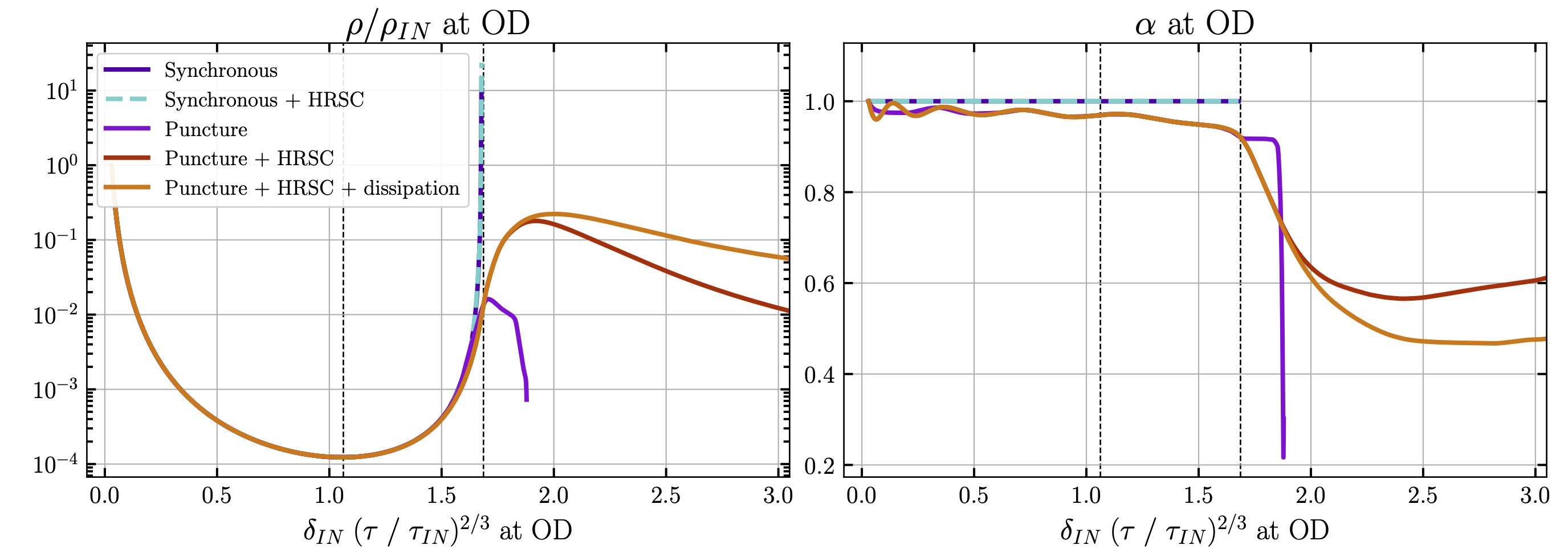}
    \caption{Normalised density (left) and lapse (right) over a rescaled proper time at the peak of the over-density. The density decreases with the expansion of the universe, until turnaround at $\delta^{(1)}_{TA} \simeq 1.062$, after which the over-density collapses inward at $\delta^{(1)}_{C} \simeq 1.686$ (both marked with vertical lines). This figure shows the impact of including HRSC and dissipation, which only help stabilise the simulation when using a puncture gauge,  \cref{eq:puncture}.}
    \label{fig:HRSC_diss}
\end{figure}

Ideally, we need to slow or freeze shell-focusing with the lapse, so the peak density is frozen in time, while the surrounding region continues to evolve and display an apparent horizon. We do see this behaviour in the ``Puncture + HRSC" red curve in \cref{fig:HRSC_diss}, where the lapse $\alpha$ decreases up to a minimum and then gradually increases back up (see explanation below \cref{eq:KmKFLRW}), but this minimum of $\sim0.55$ needs to be lower. So we need to find a stable slicing that lets the evolution proceed as far as possible and further decreases this minimum, which we do with dissipation.

Including Kreiss-Oliger artificial dissipation \cite{Kreiss_1973, Bozzola_2021} adds an extra term to all evolution equations of the simulation. As an example, consider a variable $u$ that evolves according to $\partial_t u = S$, a dissipation term is included as:
\begin{equation}
    \partial_t u = S + \epsilon \frac{dx^5}{2^{6}}(\partial^{\{6\}}_x + \partial^{\{6\}}_y + \partial^{\{6\}}_z )u,
\end{equation}
with $\partial^{\{6\}}_i$ spatial gradients calculated with $6^{th}$-order stencils.
This is the specific form of dissipation relevant for the simulations of this paper, since $dx=dy=dz$ and we use $4^{th}$-order finite-difference stencils for all our evolution equations. We can directly see that, as we explore convergence, $dx \rightarrow 0$ (see \cref{sec:Shift&MR}), the $dx$ factor will disappear at a $5^{th}$-order convergence rate. Indeed, the dominating finite-difference error of our simulations is $4^{th}$-order, so this dissipation term should converge away more quickly than the scheme's convergence toward the physical solution.

The $0 \leqslant \epsilon \leqslant 1/3$ value tunes the strength of this dissipation, where we use $\epsilon = 0.2$ constant on all levels; see \cite{Bozzola_2021} for more advanced methods. We chose this value because it strongly supports the steepening profile, and high-resolution tests with larger $\epsilon$ values would crash earlier without further decreasing the lapse.

So, while in practice this means adding non-physical terms to the evolution, their role should always be subdominant, supporting stability by damping numerical noise. As the orange curve in \cref{fig:HRSC_diss} shows, this allows the over-dense profile to steepen much further, with increased $\rho$ and decreased $\alpha$. Then, as resolution improves, these dissipation terms decrease accordingly.

\section{Lapse evolution} \label{sec:Lapse}

\begin{figure}[t]
    \centering
    \includegraphics[width=\linewidth]{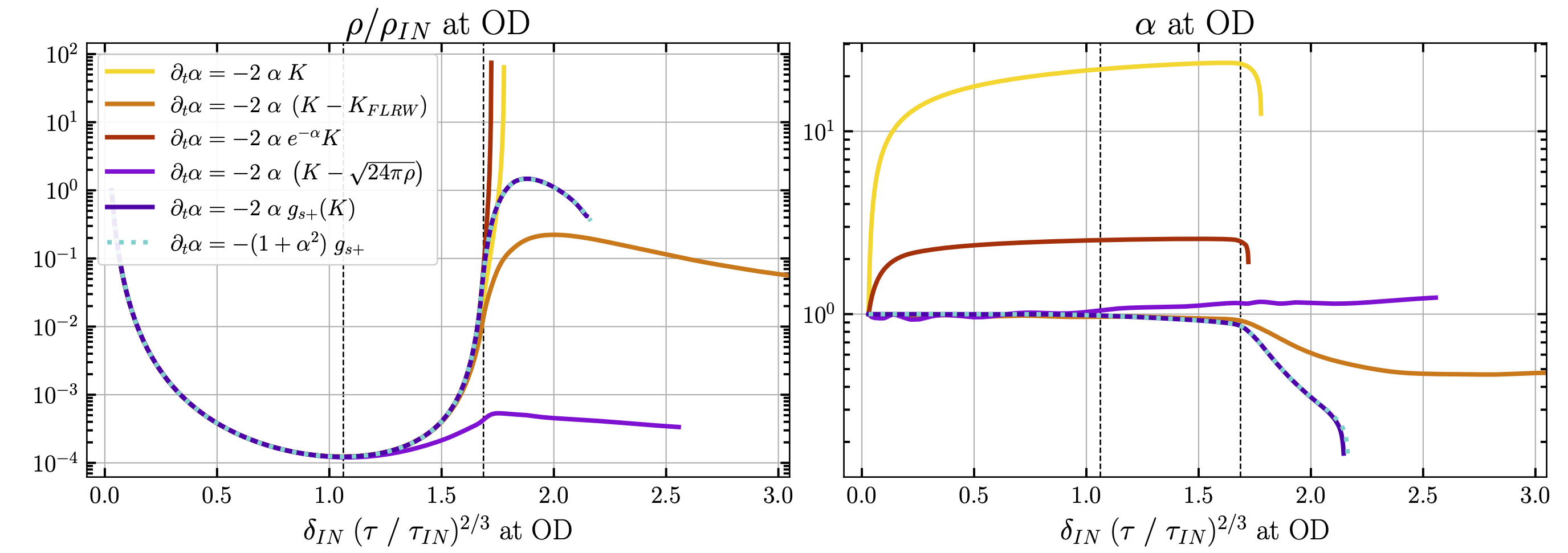}
    \caption{Normalised density (left) and lapse (right) over a rescaled proper time at the peak of the over-density. The density decreases with the expansion of the universe, until turnaround at $\delta^{(1)}_{TA} \simeq 1.062$, after which the over-density collapses inward at $\delta^{(1)}_{C} \simeq 1.686$ (both marked with vertical lines). This figure shows the impact of different lapse evolution equations while including HRSC and dissipation, and letting the shift $\beta^i$ evolve according to \cref{eq:puncture}. Note that the orange line showing $\partial_t \alpha = -2\alpha (K-K_{FLRW})$ is also present as the orange line in \cref{fig:HRSC_diss}.}
    \label{fig:lapse_evo}
\end{figure}

Here we build upon the best case of the previous section, the orange ``Puncture + HRSC + dissipation" curve in \cref{fig:HRSC_diss} (also orange in \cref{fig:lapse_evo}). Now we explore improved lapse slicing conditions to induce stability in this simulation while allowing $\alpha$ to decrease. We choose how the lapse evolves starting from the Bona-Masso \cite{Bona_1994} family of slicing conditions parametrised as:
\begin{equation}
    \partial_{t}\alpha = - \alpha^2 f_{\rm BM}(\alpha) \; g(\alpha,\;K),
\end{equation}
with $f_{\rm BM}$ an arbitrary positive function, and instead of the usual $K$, we introduce a $g$ function for the cosmological context. A common choice for BH simulations is to set $f_{\rm BM}$ to the 1+log slicing \cite{Anninos_1995} with
\begin{equation}
    f_{\rm BM}(\alpha) = 2 / \alpha,
\end{equation}
which has strong singularity avoiding properties, freezing time before its appearance\footnote{Note, that time is ``frozen" when $\alpha \rightarrow 0$, so these equations are implemented in the code as $\partial_{t}\alpha = - 2\alpha g$ and not $\partial_{t}\alpha = - \alpha^2 (2/\alpha) g$, and likewise for the shock avoiding slicing.}. Another choice with similar properties, but with the added benefit of maintaining smooth and evenly spaced coordinates, is the shock-avoiding slicing \cite{Alcubierre_1996, Baumgarte_2022}
\begin{equation}
    f_{\rm BM}(\alpha) = 1 + \frac{1}{\alpha^2}.
\end{equation}
The numerator of the second term is a free constant commonly set to one, as larger values enable stronger oscillations \cite{Baumgarte_2022}. We test and present both choices in \cref{fig:lapse_evo} as the black solid line and the cyan dotted line. These figures indicate that both $f_{\rm BM}$ options show similar evolution, but the shock-avoiding slicing allows a slightly longer evolution before the simulation crashes. We then continue our work using the shock-avoiding slicing. However, we recognise some drawbacks of this gauge; while the 1+log gauge allows $\alpha$ tending towards zero, the shock-avoiding gauge allows $\alpha=0$ and even negative values. This has been found not to pose any stability issues \cite{Baumgarte_2022}, but it challenges intuition about time perception, especially with a non-zero shift, and many matter codes commonly implement equations with $1/\alpha$ and therefore need to be adapted.

\vspace{0.5cm}

All simulations are evolved until they no longer display collapsing features, or until they crash. The first occurs when the singularity is too strongly avoided, and the second when it is unsuccessfully avoided, so we trial different $g(\alpha,\;K)$ (of dimension $1/L$) to seek a stable balance. 

Usually $g(\alpha,\;K)$ is directly set as the trace of the extrinsic curvature $g = K$ for ``standard" BH simulations. So with 1+log, $\partial_t \alpha = -2\alpha K$, this has the convenient property that when spacetime is contracting inwards $K>0$, the lapse decreases $\partial_t\alpha<0$ and so $\alpha\rightarrow 0$ and time is frozen for this coordinate grid observer. Applying this in our cosmological simulation, the result is shown in the yellow line of \cref{fig:lapse_evo}, which, however, shows a large increase in the lapse $\alpha \sim 20$ before even engaging in the singularity-avoiding behaviour. A decrease in the lapse is clearly visible towards the end, but the distance to cover $20 \rightarrow 0$ is too large for the gauge to reach in a stable manner. This is because, in the cosmological context, the extrinsic curvature is initially dominated by the expansion of the universe and so accelerates the simulation's time evolution. Even later, when structures are gravitationally bound, expanding void regions in cosmological simulations induce lapse growth instead of the collapse needed.

Therefore, several suggestions have been proposed to modify $g(\alpha,\;K)$ and account for cosmological expansion. Previous work would remove the initial $K$ value \cite{Baiotti_2004}. So naturally, a first step for cosmologists was to remove the cosmological expansion. This has been implemented with the average $K$ over the simulation box, $g = K - \langle K \rangle$ \cite{Giblin_2019a}, or alternatively with the simulation box edge value, $g = K - K_{\rm edge}$ \cite{Yoo_2013}. These methods similarly remove the homogeneous part, so the inhomogeneity determines the time evolution.

For our simulations with \texttt{Einstein Toolkit} \cite{ET2012, ET2025}, the gauge evolution equations are provided by our thorn \texttt{CosmoLapse} \cite{CosmoLapse} (other codes may call this a module; with Einstein Toolkit, it is a thorn). However, we use \texttt{Kranc} \cite{Kranc} to generate \texttt{CosmoLapse} from a Mathematica file, and so the implemented evolution equations are limited to local values, meaning we cannot use global quantities such as the average or edge. As an alternative, we use the analytical expression of the homogeneous fluid expansion:
\begin{equation} \label{eq:KmKFLRW}
    g(\alpha,\;K) = K - K_{\rm FLRW},
\end{equation}
where $K_{\rm FLRW} = -3H = - 2/\tau$ and here $\tau$ is the local value of proper time calculated in \cref{eq:tau}. Each grid position has its own $\tau$, so if the coordinates were synchronous, $K_{\rm FLRW}$ would be homogeneous on the given slice, but here, as $\partial_t\alpha\neq0$, $K_{\rm FLRW}$ is spatially varying. We present this equation's evolution as the orange curve in \cref{fig:HRSC_diss} in the previous section and also as the orange curve in \cref{fig:lapse_evo}. To describe its evolution, we identify three phases. 

The first initial phase goes up to $\delta^{(1)}_C$ and contains some oscillations. These probably come from the discrepancy between the $\partial_t\alpha=0$ initial conditions and the inhomogeneous $\partial_t\alpha$ evolution. Indeed, comparable initial spurious radiation is typically found in binary BH simulations \cite{Higginbotham_2019}. Furthermore, here the subtracted term $K_{\rm FLRW}$ evolves in time, so the evolution is regularly course-corrected as the target asymptotic limit varies. 

The second phase, after $\delta^{(1)}_C$ up to $\delta^{(1)} \simeq 2.7$, displays a drop in the lapse up to $\alpha \sim 0.5$ and a decrease in the density. This collapse of the lapse indicates time tending toward freezing; however, the lapse would need to decrease further than $0.5$. The volume at the peak increases as the shift pushes coordinate positions away, trying to stabilise this shell-focusing region and therefore decreasing the density. 

After that, in the third phase, we see the lapse gradually increasing back up, which is an issue with this gauge. Indeed, $g$ oscillates around zero, with a pronounced negative dip during the collapse. However, now in this final phase, $g$ still oscillates but around a negative value, meaning $\partial_t\alpha>0$ and $\alpha$ is lifted back up. Should a BH appear, this slicing will not be able to resolve it, so we explore further variations of $g(\alpha, \; K)$.

Turning towards other examples in the literature, we also tried \cite{Doherty_2025}
\begin{equation}
    g(\alpha,\;K) = K + \sqrt{24 \pi \rho},
\end{equation}
which is inspired by the Friedmann equation. With our initial conditions, this is in magenta in \cref{fig:lapse_evo} and shows similar initial oscillations, but it did not display lapse-collapsing behaviour before the simulation crashed. 

Rather than removing a term based on cosmic expansion, there are direct attempts to dampen the lapse growth with \cite{Elley_2024}
\begin{equation}
    g(\alpha,\;K) = e^{-\alpha}K.
\end{equation}
See its impact in our simulations in red in \cref{fig:lapse_evo}. With $\partial_t\alpha=-2\alpha K$, it reduced the lapse growth from 20 to 2, which is, however, still not enough to enable a stable collapse.

So, we attempted new functions. First, considering the matter growth, setting $\alpha\propto(1+\delta)^{-1}$ such that voids with $\delta \rightarrow 0$ would have $\alpha \rightarrow 1$ and over-densities with $\delta \rightarrow +\infty$ would have $\alpha \rightarrow 0$. Simulations initially behaved as expected; however, they crashed because of Courant–Friedrichs–Lewy (CFL) instabilities. Indeed, because the density grows exponentially in the over-density, using it in a differential equation can pose stability issues \cite{LeVeque_2002}.

We go back to considering the extrinsic curvature, appreciating the puncture-avoiding behaviour and wishing to cancel the cosmological expansion. We take a variant of $\max(0, \; K)$, so that when expanding ($K<0$) the lapse continues to be synchronous, but when contracting ($K>0$) the slicing transitions to follow the conventional puncture gauge. However, the $\max$ function introduces a discontinuity, which could introduce numerical noise when spatial derivatives are calculated. We instead consider the softplus function
\begin{equation} \label{eq:softplus}
    g_{s+}(\alpha,\;K) = \dfrac{1}{S}\ln\left(1 + e^{S (K - T)}\right) + T,
\end{equation}
which is a smoothed max function, with $T$ the transition value with the same dimensions as $K$ ($1/L$), and the dimensionless $S$ tunes the steepness of this transition. If $S$ is too large, the transition is too steep for finite difference methods to resolve the gradient. Then, if $S$ is small, the elongated gradual gradient skews the evolution as $K \rightarrow T^{-}$; this transition zone would pass on a $g_{s+}>T$. So, through trial and error, the largest value we could simulate stably was $S=450$, and now we can focus on the transition value.

Initially, $T=0$, following the synchronous gauge, but once the slicing completes its transition to follow $K$, it is important for the gauge to evolve in an uninhibited manner. We therefore update the transition value to correspondingly decrease during the simulation following
\begin{equation}
    T = - \max\left(-T, g_{s+}(\alpha,\;K)\right).
\end{equation}
This is presented in \cref{fig:lapse_evo}, with 1+log in black and shock-avoiding in cyan dotted, which is our best result of this section and upon which we build in the next sections. Indeed, $\alpha$ shows a stable evolution, and after the collapse, $\alpha$ decreases in a stable manner towards zero. However, this simulation still crashes, so we turn to the shift in the next section to stabilise it.

\section{Shift evolution and mesh refinement} \label{sec:Shift&MR}

While the main challenge in the lapse evolution equation is countering cosmological expansion, the challenge with the shift is the shell-focusing singularity. It needs to be precisely the perfect strength to move coordinate-grid observers away, such that the central region continuously represents a larger volume bound to form a shell-focusing singularity later. This shift has to endlessly postpone shell-focusing by preventing it from being resolved, while the surrounding region continues to evolve. Stable BH simulations show that, this continuous under-resolution of the high-curvature region, gives late-time slices a trumpet shape which matches a wormhole solution \cite{Hannam_2008}. This topology is also found in stellar gravitational collapse, the reason being that the shift Gamma-driver gauge stretches the numerical spatial grid pushing all matter into an under resolved region while a vaccum BH grows \cite{Thierfelder_2011}. 

With such aspirations, but applied to our dust cosmological case, we consider the standard Gamma-driver gauge \cite{Alcubierre_2008}
\begin{align*}
    \partial_{t} \beta^{i} &= B^{i},\\
    \partial_{t} B^{i} &= \tfrac{3}{4} \; \partial_{t}\tilde{\Gamma}^i  - \eta B^{i}.
\end{align*}
It is designed to transfer any distortions, as measured by the conformal contracted Christoffel symbol $\tilde{\Gamma}^i$, into shift variations. The factor in front of $\partial_{t}\tilde{\Gamma}^i$ determines the response speed; we use the standard $0.75$ value and exclude $\alpha$ so that the shift can still react when the lapse decreases. The last term is damping with the damping factor $\eta$ typically set to $2/M$, with $M$ the final mass of the BH; indeed, we use a BH singularity-avoiding gauge to avoid a shell-focusing singularity.

Throughout all simulations presented in this paper so far, $\eta=1$, which was chosen arbitrarily. The total initial mass in the simulation box is 0.47$L$, where the box size $L$ is our unit of measurement since $G=c=1$. Going forward, we shall tune $\eta$ and, for all simulations in this and the next section, the lapse evolution equation is $\partial_t\alpha = -(1 + \alpha^2)g_{s+}$ since it performed the best in \cref{sec:Lapse}.

\begin{figure}[t]
    \centering
    \includegraphics[width=0.8\linewidth]{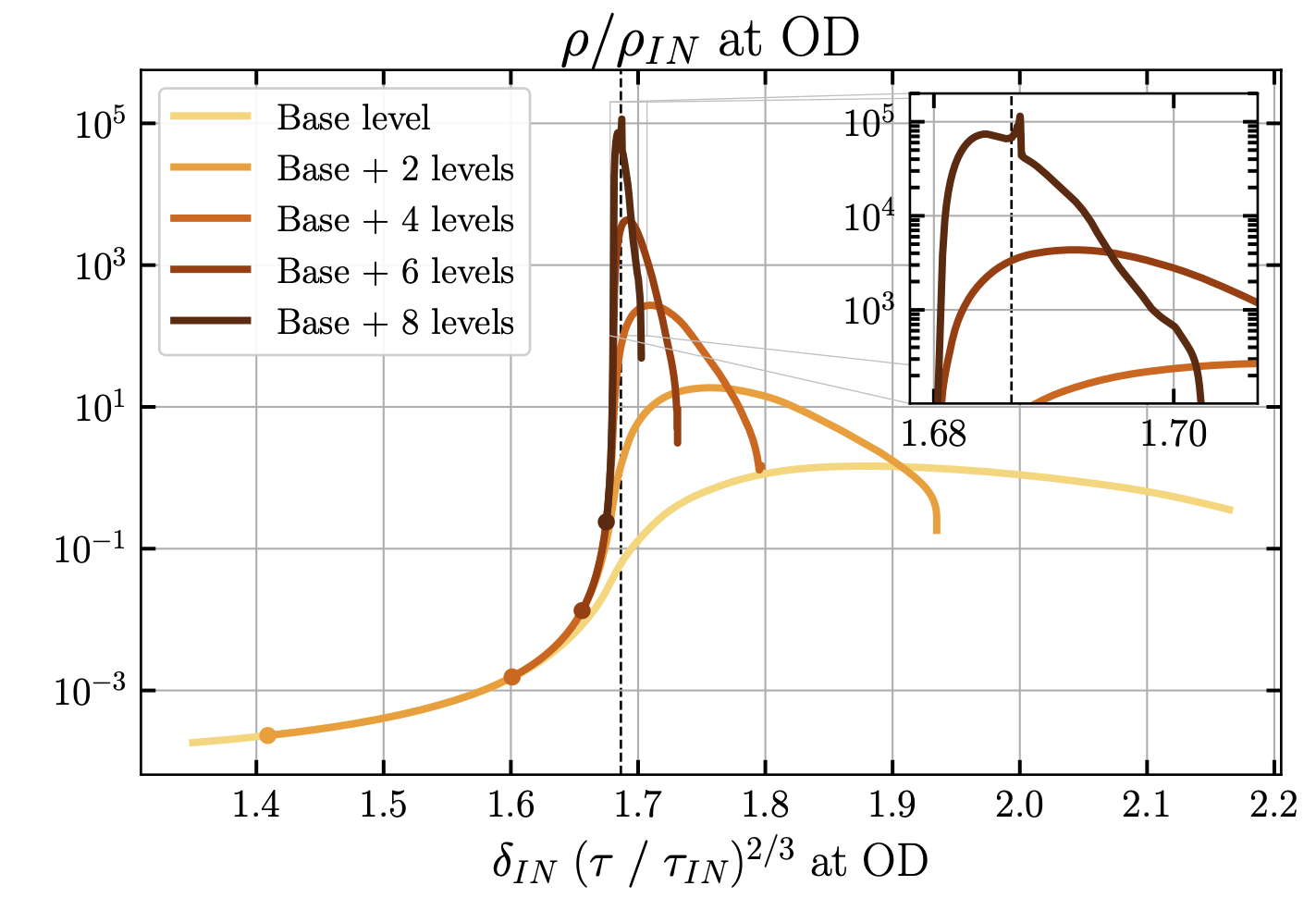}
    \caption{Normalised density over a rescaled proper time at the peak of the over-density for simulations with shift damping $\eta=1$ and increasingly larger number of mesh refinement levels. The scatter dots indicate the beginning of each higher-resolution simulation, and the vertical dashed black line marks the expected collapse time $\delta_{C}^{(1)} \simeq 1.686$. This shows that the beginning of the high-resolution simulations overlaps with the lower-resolution ones, then reaches orders-of-magnitude higher values during shell-focusing before being stabilised by the shift.}
    \label{fig:rhoMR}
\end{figure}

Typically, one determines the damping value to use through low-resolution calibration runs, then proceeds to high-resolution production runs. Indeed, all previous simulations were low-resolution with $64^3$ points. Now as we consider increasing this, we are mindful that in this cosmological context, we do not need high-resolution for the initial linear inhomogeneity, but it becomes relevant when non-linearities grow. To save needless computing resources, we developed \texttt{AddRefinements} \cite{AddRefinements} to trigger the mesh refinement thorn \texttt{CarpetRegrid2} to add more refinement levels when certain values of the proper time are reached in the central over-density. In the rescaled notion of time \cref{eq:delta1} these trigger points are at: 
\begin{equation}
    \delta^{(1)} = [1.409, \; 1.546, \; 1.601, \; 1.629, \; 1.656, \; 1.669, \; 1.675, \; 1.677].
\end{equation}
NB: from the spherical collapse model, turnaround from expansion to infall is at $\delta_{TA}^{(1)} \simeq 1.062$ and collapse is at $\delta_{C}^{(1)} \simeq 1.686$. We chose these trigger-point values so that, once a new level was included, the first couple of time-steps would overlap with the coarser level's results before steepening, see \cref{fig:rhoMR}.

The first notable feature from \cref{fig:rhoMR} is that the central density value goes multiple orders of magnitude higher as resolution is refined. This indicates that the refinement zooms in on the divergence arising from the shell-focusing singularity. After the density growth, the shift decreases this central $\rho$ value to stabilise the simulation. However, because higher-resolution simulations crash earlier, the shift seems less effective.

\begin{figure}[!ht]
    \centering
    \includegraphics[width=0.85\linewidth]{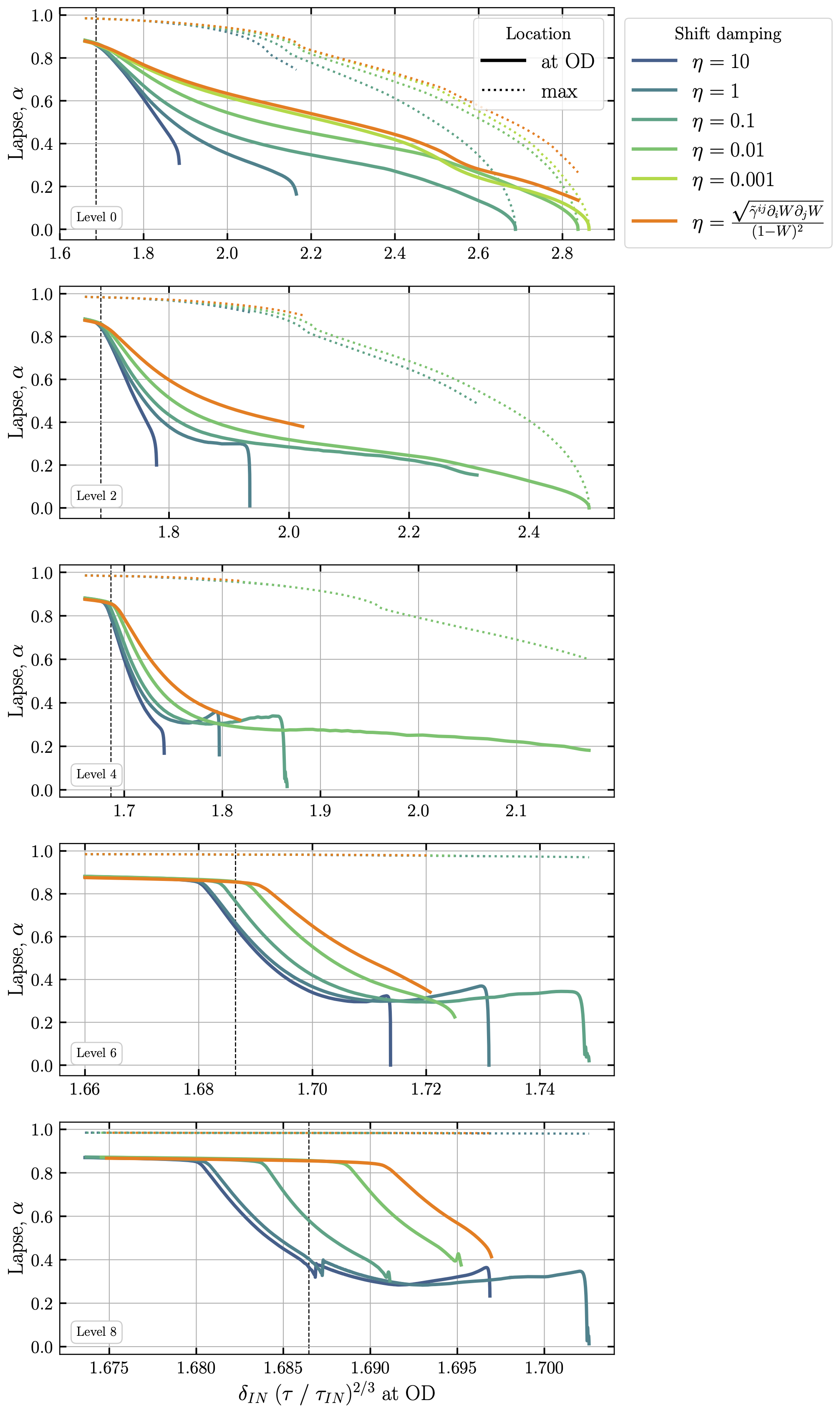}
    \caption{Lapse at the peak of the over-density (full lines) and its maximum (dotted lines) during simulations with differing damping values in the shift evolution equation. Each panel represents different sets of simulations with different amounts of mesh refinement; Level 0 is for base-level coarse simulations, Level 2 has simulations with base-resolution plus two mesh refinement levels centred on the peak of the over-density, and so on for Level 4 and up. The horizontal axis is over a rescaled proper time (function of the lapse) where the vertical dashed black line marks the expected collapse time, $\delta^{(1)} \simeq 1.686$ according to the spherical collapse model.}
    \label{fig:shiftMR}
\end{figure}

To explore this further, see \cref{fig:shiftMR} showing the lapse's evolution in simulations with different damping values and refinement levels. First considering the coarsest base level `Level 0', we see that the weaker the damping, the longer the simulation lasts. At higher refinement levels, the lapse decreases more steeply (as the horizontal-axis range shrinks), and the best-performing damping value is resolution-dependent. Indeed, the damping value that allows the simulation to last the longest is different for different refinement levels: $\eta=1$ for Level 8, $\eta=0.1$ for Level 6, $\eta=0.01$ for Level 4… This suggests that $\eta$ should not be fixed across all levels, but should instead respond to the profile's steepness as revealed by different levels. No matter the level of refinement, if the gauge is inappropriate, it will not be possible to have a stable BH simulation \cite{Snajdr_2006}.

Previous discussions warn of numerical instabilities for large damping on coarse levels \cite{Schnetter_2010}\footnote{As this instability depends on the refinement level configuration, we note that our simulations use sub-cycling for all levels with buffer zones between the refinement boundaries.} and so suggests tapering $\eta$ down at large radii, even evolving it with its own evolution equation \cite{Alic_2010}. In our case, we do not encounter these CFL-like instabilities, as they are not present in the coarse simulations; see Level 0 in \cref{fig:shiftMR}. Nevertheless, we do try one of the proposed varying damping values depending on the $W$ BSSNOK conformal factor \cite{Muller_2010a, Muller_2010b}:
\begin{equation}
    \eta = \frac{\sqrt{\tilde{\gamma}^{ij}\partial_i W \partial_j W}}{(1-W)^2}.
\end{equation}
This is the orange line in each plot of \cref{fig:shiftMR}, showing a slower lapse compared to all other $\eta$ values tested, and an earlier simulation end than some of the other options. This may be due to the inclusion of additional derivatives in the simulation's evolution, how they are implemented with \texttt{CosmoLapse} and \texttt{Kranc}, and $\eta$ should also be tuned by adjusting an overall factor in front of this expression (set to $1$ in our attempt here).

However, in all cases, the simulation still crashes. On the coarser levels, the simulations stop as they reach an issue with the lapse evolution equation. This was explained after \cref{eq:softplus}; at large radii the extrinsic curvature that started negative with the expansion of the universe tends towards zero, $K \rightarrow 0^{-}$ entering the transition zone of the softplus function where $g_{s+}>0$, meaning $\partial_t\alpha = -(1+\alpha^2) g_{s+} < 0$. Therefore, even the void region will have a decreasing lapse. This is shown by the maximum lapse value in the simulation box, represented as the dotted lines in \cref{fig:shiftMR}. When the maximum reaches $0$, the whole simulation box stops evolving, and so the simulation stops.

Meanwhile, at high-resolution this is not an issue; indeed, these dotted lines remain around $1$ throughout the simulations. Instead, those simulations crash because of numerical instabilities that arise with these large gradients.

A quick note on advection: in this paper we deliberately avoided mentioning advection terms in the lapse or shift evolution equations (such as $\beta^k\partial_k\alpha$, $\beta^k\partial_k\beta^i$, $\beta^k\partial_k\tilde{\Gamma}^i$ and $\beta^k\partial_k B^i$). This is because whether these added terms support or not the stability of these simulations also seems to depend on resolution. It seems advection terms in the shift evolution make no difference at low resolution, and make simulations crash earlier at high resolution. Advection terms in the lapse make the simulation crash earlier at low resolution but last longer at high resolution, and even more so if the finite difference scheme used is upwinded\footnote{Classic finite difference schemes centre the stencil around the point of interest, calculating its gradient based on left and right neighbouring points. Upwinded schemes place the stencil on one side, considering only the left or right neighbouring points. This is usually preferred when calculating a gradient contracted with a vector $v^i\partial_i$, so that only points in the vector's `past' are considered.}. These effects, however, are marginal compared to the large differences observed with the damping.

We've shown here that the usual approach to balancing the numerical resolution and the gauge induced under resolution does not work in this dust cosmological context. The resolution competition is balancing out differently for this shell-focusing singularity compared to what is expected for a BH singularity. So more work is needed to find an appropriate shift evolution equation for shell-focusing singularities. This could be informed by constructing local Penrose diagrams \cite{Hannam_2008}.

\section{Black hole criteria} \label{sec:BHcrit}

Nevertheless, some simulations in \cref{fig:shiftMR} show interesting features which are best exemplified by the $\eta=0.1$ Level $6$ case. As we seek a gauge that postpones shell-focusing long enough for the surrounding environment to show an apparent horizon indicative of a BH, we monitor for this using two diagnostics: Schwarzschild radius with the Hoop conjecture in \cref{sec:Hoop} and outgoing null ray expansion in \cref{sec:nullexp}. Whether an apparent horizon is present or not, there is a growing singularity that we also study considering the contributions to the Kretschmann scalar in \cref{sec:RW}. All results in this section have been calculated with our numerical analysis Python package \texttt{aurel} \cite{aurel}. This code can calculate any local relativistic quantity by automatically resolving computational pathways based on available data. This makes it highly flexible, reliable and easily extendable.

The distribution of matter in this simulation \cref{fig:ICdensity} is not spherically symmetric, especially as this shape sharpens throughout the simulation. We determine three directions of interest starting from the peak of the over-density at the origin $x=y=0$:
\begin{itemize}
    \item \textit{vertex} along the $x$-direction with $y=z=0$, which is always over-dense
    \item \textit{edge} along $x=y$ direction with $z=0$, which is over-dense then under-dense
    \item \textit{face} along the $x=y=z$ direction, which likewise is over-dense then under-dense and goes through to the centre of the under-dense void.
\end{itemize}
This naming is inspired by fitting an octahedron around the origin \cite{Munoz_2023}, with each coordinate direction passing through the octahedron's vertices. This shape has 6 vertices, 12 edges, and 8 faces; correspondingly, the profile along each radius direction listed above is repeated as many times about the origin.

\subsection{Hoop conjecture} \label{sec:Hoop}

The first diagnostic is considering the Schwarzschild radius with the hoop conjecture \cite{Thorne_1972}. This mainly states that a BH is present when a circular hoop of circumference $4\pi M$ can be placed and rotated around the imploding object, with $M$ the mass within. 

For this, we ought to first define the \emph{coordinate radius} $r$. This starts at the origin $x=y=z=0$ (the grid cell centre) and extends to the grid cell centres throughout. It is simply calculated from the Cartesian coordinates $r = \sqrt{ \delta^{ij} x_{i} x_{j} }$ or, more simply, by summing the $dr$ elements. Depending on the direction $dr$ is rescaled as: vertex $dx$, edge $\sqrt{2}dx$, and face $\sqrt{3}dx$. On several occasions, we use $\tfrac{1}{2}dr$ for the grid cell at the origin; this pushes the endpoint of the radial measurement onto the grid cell edges rather than the centres. For example, in the central grid cell at the origin, along the vertex direction, the coordinate radius goes from the grid cell centre to the grid cell edge: $\tfrac{1}{2}dr$. Thus, we defined the coordinate radius.

In general relativity, distances are dynamical and measured according to the evolving spatial metric. So for each direction, the \emph{proper radius} in the $i^{th}$ grid cell is:
\begin{equation} \label{eq:proper_radius}
    R_i = \sum_{j=1\rightarrow i} \left(\gamma_{j}^{1/6} \;dr\right) + \gamma_{0}^{1/6} \;\tfrac{1}{2}dr,
\end{equation}
with $\gamma_j$ the determinant of the spatial metric in the $j^{th}$ grid cell. This is measured from the centre of the grid cell at the origin to the edge of the $j^{th}$ grid cell. 

At this point, we start simplifying the calculation by assuming spherical symmetry. We have calculated the proper radius along each direction; now we seek the proper volume. To truly calculate this, we ought to integrate over $\gamma$ in all possible directions, polar $0\rightarrow\pi$ and azimuthal $0 \rightarrow 2 \pi$. For simplicity, we calculate the proper volume as $\tfrac{4\pi}{3} R_{i}^3$ for each direction, giving three different volume estimates, and we acknowledge that the true value lies somewhere between the most (vertex) and least (face) dense directions. We take this same logic for the mass and circumference: we assume spherical symmetry, but we repeat these calculations applying them to each of the three directions to appreciate the full distribution.

Next, to determine the \emph{Schwarzschild radius} $2M$, we need the mass $M$. We cumulatively integrate the successive spherically symmetric shells:
\begin{equation}
    M_i = M_{i-1} + \rho_{i}\frac{4\pi}{3}
    \left( R_{i}^3 - R_{i-1}^3 \right).
\end{equation}
The vertex direction overestimates the mass, and the face direction starts out with a bigger mass (same density but $3^{3/2}$ larger volume) but then underestimates it further out as it reaches the under-density.

To go from the radius to the circumference, the typical Cartesian expression is $c=2\pi r$, but this no longer holds in curved space. Distances are rescaled with $\gamma^{1/6}$, so the proper circumference is $C_{\rm prop} = \gamma^{1/6} 2 \pi r$ (assuming $\gamma$ is spherically symmetric for each direction). From this, we define the \emph{proper circumferential radius} $\Re = C_{\rm prop} / 2\pi$ so
\begin{equation}
    \Re = \gamma^{1/6} \left( r + \tfrac{1}{2}dr \right),
\end{equation}
which differs from the proper radius of \cref{eq:proper_radius}. This last radius is the least intuitive, but it is necessary in general relativity. Indeed, consider a BH with a singularity in its centre; measuring the radius out from the centre is ill-defined. Instead, we measure the circumference and deduce the circumferential radius. In our simulations, as the shift increases, the volume element of the central few grid points increases, which makes $\Re$ non-monotonic.

\vspace{0.5cm}

\begin{figure}[t]
    \centering
    \includegraphics[width=\linewidth]{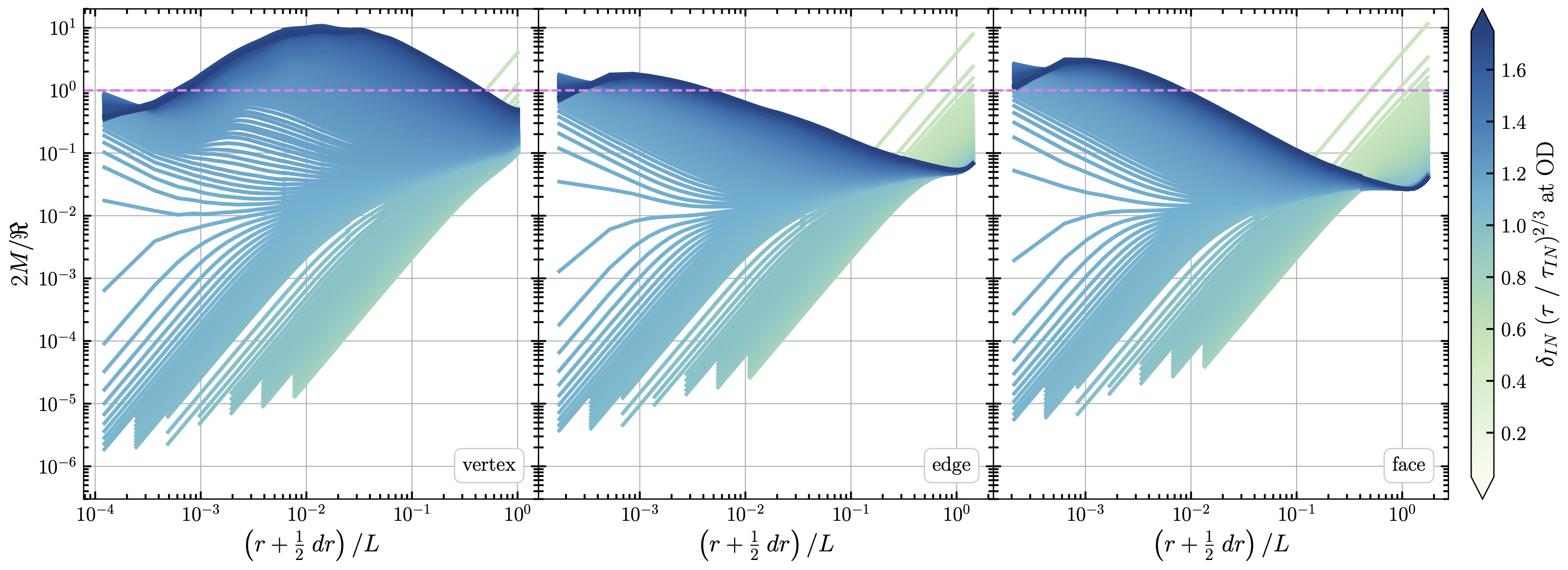}
    \caption{Ratio of Schwarzschild radius $2M$ and circumferential radius $\Re$ in each direction of the simulation box, with the coordinate radius $r$ on the horizontal axis. Each curve is a snapshot of the simulation, with colour indicating rescaled proper time, and higher resolution is included gradually during the simulation. These calculations are applied to the $\eta=0.1$ Level 6 simulation of \cref{fig:shiftMR}. This shows the Schwarzschild radius $2M$ growing throughout the simulation, relative to the circumferential radius $\Re$, as the density grows. When $2M > \Re$ (see the dashed horizontal line), the Schwarzschild radius is larger than that sphere and so would be indicative of a BH according to the hoop conjecture.}
    \label{fig:Schwarzschild_radius}
\end{figure}

Going back to the hoop conjecture, we now have all the tools to see if the Schwarzschild radius $2M$ is greater than the circumferential radius $\Re$; in which case we would expect a BH horizon to encompass the object. See the ratio of the two in \cref{fig:Schwarzschild_radius}. Each direction has a given panel, where the $2M/\Re$ ratio is calculated at different snapshots in the simulation. Lighter colours indicate earlier times in the simulation, and darker colours indicate later times. 

At early times (lighter colour), $2M/\Re$ is linear. Both $2M$ and $\Re$ grow as we go out from the centre, with $2M$ increasing more quickly. Indeed, initially at early times, the volume is quite compact, even though the cosmological horizon is smaller than the simulation box size. After this, cosmological expansion grows $\Re$ significantly, so $2M/\Re$ decreases; see the rightmost side of the face panel.

As the simulation evolves, looking in the central region, there are `jumps' between snapshots of the smallest coordinate radius. There are 6 of these because this simulation introduces 6 mesh refinement levels.

Interestingly, later snapshots reach $2M>\Re$, indicative of a BH. The $2M/\Re$ curve exceeds 1 at different snapshot times, depending on the direction. Consider the rightmost intersection where $2M/\Re = 1$ as a ``black hole radius" (we use quotation marks because this is not a sufficient indicator that a BH is actually present). We see this ``black hole radius" grow throughout the simulation as more matter is encompassed. Then, at the latest times, the $2M/\Re$ value at the central grid point decreases, as the shift increases the central volume. 

Note that in the vertex direction, the central grid points are always under one, $2M/\Re<1$. We recall that while this direction starts at the origin with the others with the same density, it has a smaller volume. Integrating further out, this direction is then always over-dense. So it seems like it would predict a BH forming around the origin and then continuing to grow, rather than a BH forming at the origin and then growing.

\subsection{Null ray expansion} \label{sec:nullexp}

The previous calculation is somewhat `back of the envelope', and insufficient to ensure an apparent horizon is present. We take this study one step further by calculating the expansion of outgoing null vectors $\Theta_{\rm out}$. Should light rays be trapped, $\Theta_{\rm out}<0$; otherwise, when $\Theta_{\rm out}>0$, photons are freely moving. The outermost surface where $\Theta_{\rm out}=0$ corresponds to the apparent horizon; this is the marker of a BH. It differs from an event horizon, which is future-informed, as it anticipates that incoming matter will never escape. But if the BH is static, the apparent and event horizons are the same surface.

To calculate this, we start by defining a radially outward-pointing space-like unit vector\footnote{To properly calculate an apparent horizon, it should be a space-like unit vector orthogonal to the apparent horizon surface (pointing out). As we do not have an apparent horizon, a radially outward-pointing vector is sufficient, and with the symmetry built into our distribution, the vectors in each direction considered may be a good approximation.}:
\begin{equation}
    s^i = \frac{\gamma^{ij}\partial_{j}r}{\sqrt{\gamma^{kl}\partial_{k}r\partial_{l}r}}.
\end{equation}
This space-like vector is combined with a time-like vector to obtain a null vector representative of outward-pointing light rays. The numerical relativity formalism of our simulation has spatial hypersurfaces $\gamma_{ij}$ foliated along its orthonormal time-like vector. So we use this time-like vector, and so have the outgoing null ray expansion \cite{Alcubierre_2008}
\begin{equation}
    \Theta_{\rm out} = \mathcal{D}_{i}s^{i} + K_{ij}s^{i}s^{j} - K,
\end{equation}
with $\mathcal{D}_{i}$ the covariant derivative with respects to the spatial metric $\gamma_{ij}$. To set expectations, for a Minkowski spacetime, this is $\Theta_{\rm out} = \tfrac{2}{r}$, while it is larger for an FLRW spacetime, which depends on time as $\Theta_{\rm out} = \tfrac{2}{ar} + 2H$. 

\begin{figure}[t]
    \centering
    \includegraphics[width=\linewidth]{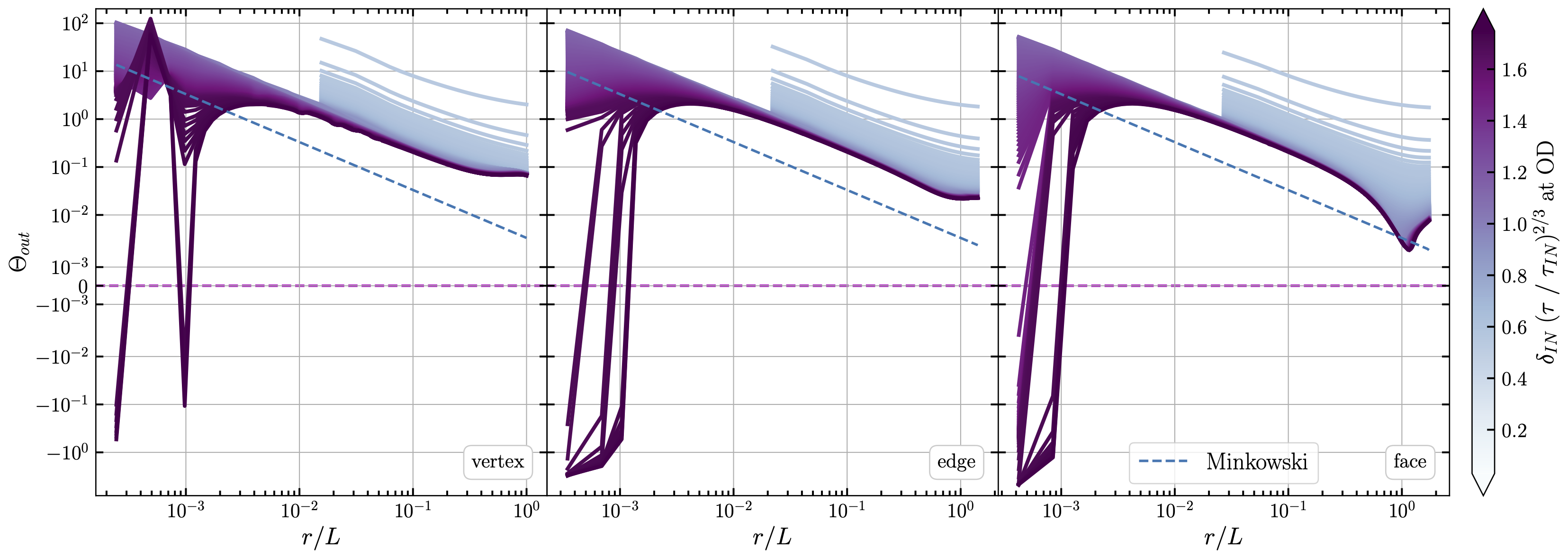}
    \caption{Expansion of a radially symmetric outward-pointing null ray unit vector as a function of the coordinate radius (with the first point cut off) in each direction of the simulation box. Each curve is a snapshot of the simulation, corresponding to the data-recording frequency, not individual time steps. The colour indicates time via the rescaled proper time, with higher resolution included gradually during the simulation. These calculations are applied to the $\eta=0.1$ Level 6 simulation of \cref{fig:shiftMR}. When $\Theta_{\rm out}<0$, photons are trapped; otherwise, they are freely moving, so the $\Theta_{\rm out}=0$ line is marked with a dashed horizontal line, and symmetric log scaling is used on the $y$-axis. 
    The blue dashed line marks the Minkowski space expectation, which is lower than the cosmological one.}
    \label{fig:Null_expansion}
\end{figure}

Results in our simulation, calculated with \texttt{aurel}, are shown in \cref{fig:Null_expansion}. The first couple of snapshots have $\Theta_{\rm out}$ decreasing with the expansion of the universe. We also note an odd feature in the face direction: the null ray expansion dips as the void is approached. We associate that with some level of focusing before entering the central region of a void; see \cite{Umeh_2014} for inhomogeneous cosmological effects.

The more interesting feature happens for late snapshots: in the central region, a couple of grid points show $\Theta_{\rm out}<0$, indicating a trapped region. However, there are not enough grid points for an apparent horizon to be numerically resolvable; indeed, \texttt{AHDirectFinder} does not find any. This can't be confirmed either by simply adding more resolution. As has been discussed in \cref{sec:Shift&MR}, more mesh refinement is not the solution; first, a more appropriate shift damping should be developed.
However, having points with $\Theta_{\rm out}<0$ is encouraging and should be pursued.

\subsection{Ricci and Weyl relative contributions to the Kretschmann invariant}\label{sec:RW}

It is interesting to consider how close we are to the formation of an apparent horizon by comparing our results with those of the spherically symmetric collapse from similar initial conditions, using invariant quantities as computed with \texttt{aurel} \cite{aurel}. Gravitational dust collapse and BH formation from cosmological spherically symmetric perturbations have recently been considered in \cite{Galoppo_2026}, starting from initial conditions based on $\mathcal{R}_c$ and essentially using the same setup explained here in Section \ref{sec:IC}.

\begin{figure}[!ht]
    \centering
    \includegraphics[width=\linewidth]{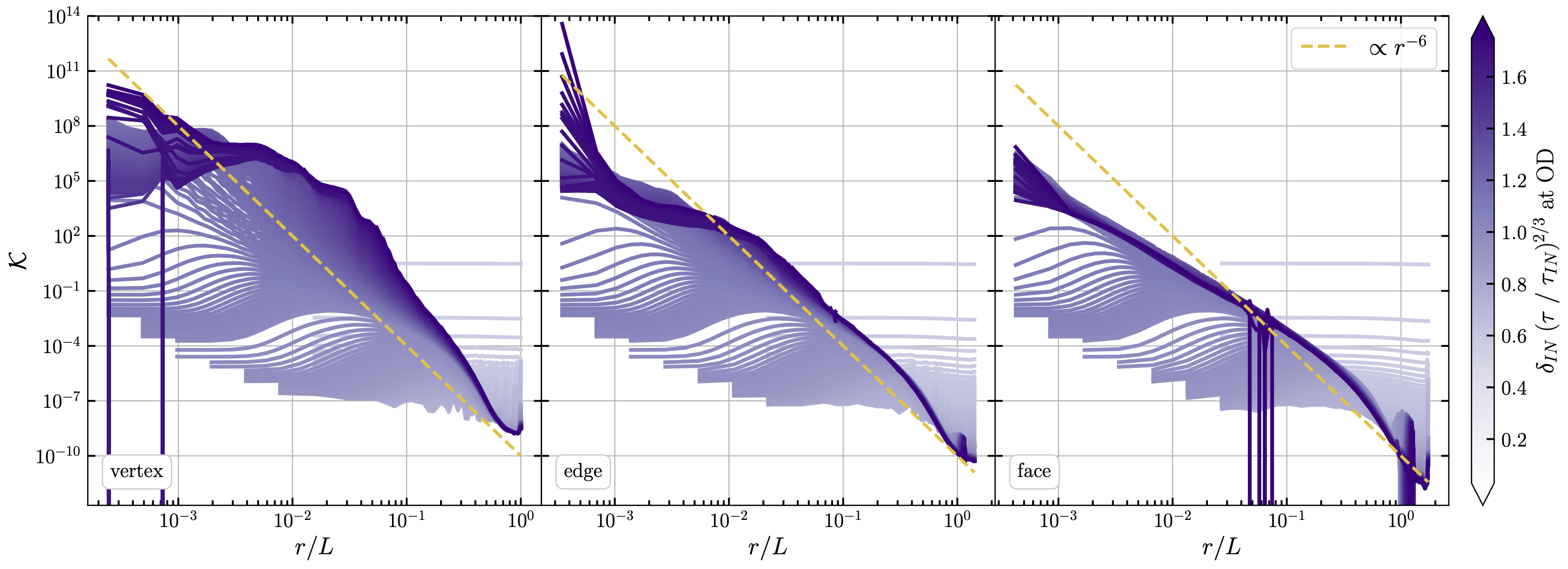}
    \caption{The Kretschmann scalar \eqref{eq:K} as a function of the coordinate radius in each direction of the simulation box. Each curve is a snapshot of the simulation, with colour indicating rescaled proper time, with higher resolution included gradually during the simulation. These calculations are applied to the $\eta=0.1$ Level 6 simulation of \cref{fig:shiftMR}. The $\propto r^{-6}$ dashed line is used as a comparison to the Schwarzschild expectation of $48 M^2 / r^6$.}
    \label{fig:Kretschmann}
\end{figure}

In its evolution, even without identifying the formation of an apparent horizon, we can expect that - at least close to the peak of the overdensity - the evolution of kinematical and dynamical quantities, and invariants built from these, will evolve similarly to the Lemaître–Tolman–Bondi (LTB) spherical and Szekeres (with a dipole) cases \cite{Galoppo_2026}. In these exact analytical models, a BH and a horizon can be clearly identified during the BH formation, and kinematical and dynamical quantities (normalised with the matter expansion) evolve to specific finite values \cite{Wainwright_Ellis_1997}. Naturally, the first invariant considered is the Kretschmann invariant scalar constructed from the full spacetime 4-Riemann tensor
\begin{equation}\label{eq:K}
    \mathcal{K} = {}^{(4)}R_{\alpha\beta\mu\nu}{}^{(4)}R^{\alpha\beta\mu\nu}.
\end{equation}
As a singularity appears in the spacetime, this invariant gives certainty it is a physical effect. In \cref{fig:Kretschmann} we show a series of snapshots of the Kretschmann invariant as a function of the coordinate radius in each direction of the simulation box. Overall, we may say that this invariant scalar function is evolving toward its  Schwarzschild counterpart, i.e.\ $\mathcal{K}=48 M^2 / r^6$ (the yellow dashed line in the figure), even if to different degrees in the three directions.

A clear transition as a BH forms is also expected in invariants from the decomposition of the Riemann tensor \cite{EBWeyl}
\begin{equation} \label{eq:Weyl}
  {}^{(4)}R_{\alpha\beta\mu\nu} = C_{\alpha\beta\mu\nu} +\left(g_{\alpha[\mu}{}^{(4)}R_{\nu]\beta} - g_{\beta[\mu}{}^{(4)}R_{\nu]\alpha}\right) - \frac{1}{3}g_{\alpha[\mu}g_{\nu]\beta}{}^{(4)}R,
\end{equation}
with its trace the Ricci tensor ${}^{(4)}R_{\alpha\beta}$ and traceless part the Weyl tensor $C_{\alpha\beta\mu\nu}$. Interestingly, \cite{Galoppo_2026} have shown that in the initial phases of the collapse, the curvature (and therefore the geodesic deviation) are Ricci-dominated (i.e. dominated by matter as in the Newtonian case), while they are Weyl-dominated at later stages. In the end, the matter contribution to collapse becomes negligible. After all, if an apparent horizon forms, surrounded by an underdensity that becomes more and more empty, the BH should asymptotically become a Schwarzschild one (or Kerr if some angular momentum was included), i.e.\ a vacuum solution of Einstein's equations $R_{\mu\nu}=0$. 

Quantifying this in the simulation, we use the trace decomposition of the Riemann tensor for the Kretschmann invariant
\begin{equation}
    \mathcal{K} = \big(C_{\mu\nu\rho\sigma}C^{\mu\nu\rho\sigma}\big) + \big(2{}^{(4)}R_{\mu\nu}{}^{(4)}R^{\mu\nu} - \tfrac{1}{3} {{}^{(4)}R}^2\big),
\end{equation}
identifying the Weyl and Ricci parts. While they both blow up in the collapse, the ratios
\begin{equation}\label{fig:Kcontributios}
\frac{C_{\mu\nu\rho\sigma}C^{\mu\nu\rho\sigma}}{\mathcal{K}}, 
~~~~
\frac{2{}^{(4)}R_{\mu\nu}{}^{(4)}R^{\mu\nu} - \frac{1}{3} {{}^{(4)}R}^2}{\mathcal{K}},
\end{equation}
remain finite; and the Weyl part should dominate, so that the Weyl ratio should be non-zero, while the Ricci ratio should become negligible. Furthermore, splitting the Weyl part into its electric and magnetic contributions (see Fig.\ 11 in \cite{Munoz_2023}), the magnetic part is negligible toward the peak (we do not have angular momentum), while the electric part dominates the geodesic deviation, leading to ``spaghettification" toward a singularity and BH formation, as detailed in \cite{Galoppo_2026}. 

\begin{figure}[!ht]
    \centering
    \includegraphics[width=\linewidth]{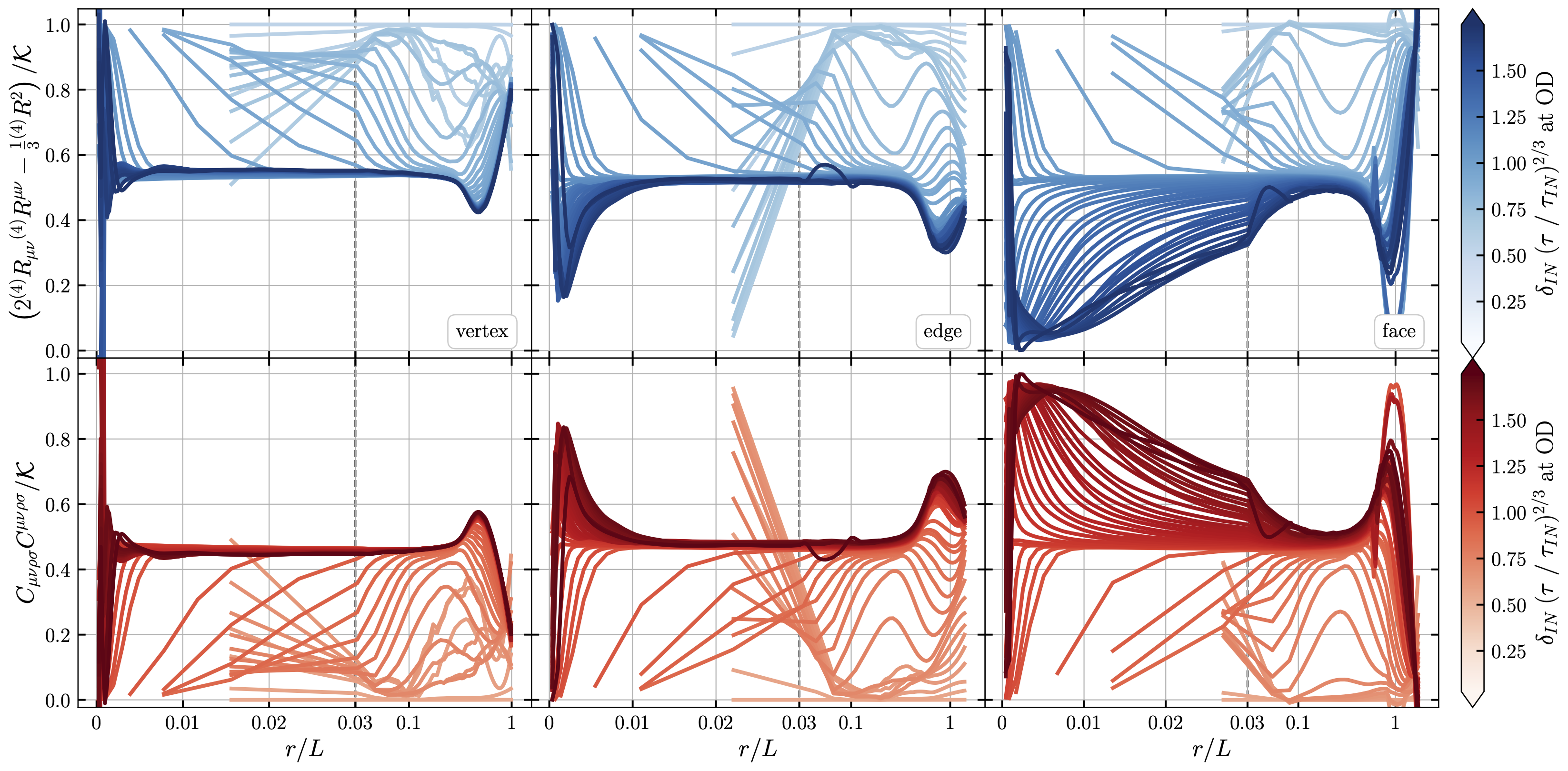}
    \caption{The Ricci (top panel) and Weyl (bottom panel) relative contributions to the Kretschmann scalar as a function of the coordinate radius in each direction of the simulation box. Each curve is a snapshot of the simulation, with colour indicating time via the rescaled proper time; higher resolution is included gradually during the simulation. These calculations are applied to the $\eta=0.1$ Level 6 simulation of \cref{fig:shiftMR}. The vertical dashed line at $r/L=0.03$ separates two types of $x$-axis scaling: linear for $r/L<0.03$ and logarithmic for $r/L>0.03$, highlighting the region where Weyl dominates.}
    \label{fig:Ricci_vs_Weyl}
\end{figure}

In \cref{fig:Ricci_vs_Weyl} we plot the relative Ricci and Weyl contributions to $\mathcal{K}$, \cref{fig:Kcontributios}, again at different times and along the three directions of the simulation box. Time frames in this figure can be compared directly with  Fig.\ 7  in \cite{Galoppo_2026}, especially the two top panels (as the two bottom panels are at times when the apparent horizon has already started to form). 


Initially, the Weyl contribution is subdominant in all three directions, while the Ricci contribution dominates. As the overdensity evolves and becomes more concentrated, the Kretschmann scalar is more equally divided between the Ricci and Weyl contributions. Towards the end of the simulation (dark coloured snapshots) near the peak, the Ricci contribution becomes subdominant while the Weyl contribution grows. This behaviour is particularly evident in the face direction, where it is well spread across the simulation box, while it is extremely concentrated in the vertex direction. However, at very small $r/L$, even $r=0$, we can clearly see that the Weyl contribution is totally subdominant. This behaviour is qualitatively similar to the one in the top-right panel of Fig.\ 7 in \cite{Galoppo_2026}, representing the time when the apparent horizon starts to form at $r=0$: we actually reach a higher Weyl contribution at the peak of the distributions in the three directions, yet these figures show that overall we are still Ricci-dominated at very small $r/L$, qualitatively (but invariantly) indicating that we have not yet entered the Weyl-dominated phase that characterises the central region once the apparent horizon has started to form, as we expect from the analysis of the spherical case in \cite{Galoppo_2026}.

\section{Conclusion} \label{sec:Conclusion}
To conclude, much more work is needed to enable full exploration of PBH formation during a matter-dominated era with numerical relativity. The challenge spans multiple fronts: for cosmological gauges and for shell-focusing singularities. Our contribution to addressing these questions is in development (with the codes we make publicly available), in insights gained on the problem (as we present how it reacts to different gauges and numerical schemes), in the proposal of a new lapse evolution equation, and in highlighting the relationship between the shift and refinement level.

For this work to be possible, we have developed and made public numerous thorns to work with \texttt{Einstein Toolkit}: \texttt{ICPertFLRW} \cite{ICPertFLRW, ICPertFLRW_GRH} to set cosmological initial conditions, \texttt{CosmoLapse} \cite{CosmoLapse} to pick and use different gauges relevant for cosmology, and \texttt{AddRefinements} \cite{AddRefinements} to gradually add refinement levels based on cosmological triggers. For analysis, we developed the Python package \texttt{aurel} \cite{aurel, EBWeyl}; this is independent of the numerical relativity evolution code and of the physical scenario considered.

Using these tools, we simulated cosmological dust collapse. We showed that these simulations need high-resolution shock capturing and dissipation to evolve stably. Testing various lapse slicings, where the challenge is that the cosmological expansion is present in the extrinsic curvature and so cannot be directly used in the same way as it is used in ``classic'' relativistic collapse simulations. We proposed a new evolution equation, based on the softplus function, that allows a transition between two regimes in the simulations, from synchronous to puncture. While this gave better chances of seeing an apparent horizon appear, this slicing needs further improvement.

Going to the shift, we show that the challenge there is related to the level of refinement. Indeed, the size of the shell-focusing singularity is determined by the grid size, so the shift damping appropriate for one level of refinement is inappropriate for another. We find that spatially varying damping is more suitable for this problem, but further work is necessary to identify which damping to use. With the best-performing simulation, we calculate BH diagnostics. These hint at a potential apparent horizon forming; however, the simulation's stability does not allow it to run long enough to resolve this. 

Interesting future directions include further adapting the softplus threshold in the lapse slicing and considering varying damping terms for the shift, which will be required for a shell-focusing singularity-avoiding gauge. Then, of course, there is the potential to avoid the shell-focusing limitation altogether by considering different initial conditions that form an apparent horizon before shell-focusing, where \cite{Yoo_2026} take the amplitude-dependent perspective and \cite{Galoppo_2026} the shape-dependent one.

\section{Acknowledgements}

We are grateful to Daiki Saito and Kazuya Koyama for useful conversations. We appreciate the support from Rapha{\"e}l Kou and Jaime Calder{\'o}n-Figueroa, and from the support teams of the different high-performance computing (HPC) clusters used.

RM is supported by STFC grant ST/X001040/1. CB is supported by STFC grants ST/X001040/1 and ST/X000796/1.

Simulations presented in this paper were run on the Artemis HPC, supported by the University of Sussex, and the Cosma8 DiRAC HPC facility, following a RAC18 allocation for this project. The DiRAC@Durham facility is managed by the Institute for Computational Cosmology on behalf of the STFC DiRAC HPC Facility (www.dirac.ac.uk). The equipment was funded by BEIS capital funding via STFC capital grants ST/P002293/1, ST/R002371/1 and ST/S002502/1, Durham University and STFC operations grant ST/R000832/1. DiRAC is part of the National e-Infrastructure.

We also acknowledge the use of the Apollo HPC, supported by the University of Sussex (Artemis' predecessor), and the Sciama HPC cluster, supported by the Institute of Cosmology and Gravitation at the University of Portsmouth.

All codes are publicly available as referenced, and simulation descriptive and derived data (including everything used to make the plots in this paper and more) are publicly available at the University of Sussex Figshare, \textsc{doi}: \href{https://doi.org/10.25377/sussex.33593488}{10.25377/sussex.33593488}.

For the purpose of open access, we have applied a Creative Commons Attribution (CC BY) licence to any Author Accepted Manuscript version arising.

\vspace{0.5mm}

In memory of Hayley.

\printbibliography
\end{document}